\documentclass[aps,prl,twocolumn]{revtex4-1}
\usepackage{amssymb}

\usepackage{graphics}
\usepackage{dcolumn}
\usepackage{pifont}
\usepackage{graphicx}
\usepackage{amsmath}
\usepackage{verbatim}
\usepackage{epstopdf}
\usepackage{amssymb}
\usepackage{color}

\newcommand{\bom}{\mathbf}
\newcommand{\tm}{\mathrm}

\newcommand{\bg}{\boldsymbol}

\newcommand{\beq}{\begin{equation}}
\newcommand{\eeq}{\end{equation}}
\newcommand{\bea}{\begin{eqnarray}}
\newcommand{\eea}{\end{eqnarray}}
\newcommand{\bwt}{\begin{widetext}}
\newcommand{\ewt}{\end{widetext}}

\begin{document}
\title{Chiral Spin Waves in Fermi Liquids with Spin-Orbit Coupling}

\author{Ali Ashrafi
and Dmitrii L. Maslov
 }

\begin{abstract}
We predict the existence of chiral spin waves--collective modes in a two-dimensional Fermi liquid 
 with the  Rashba or Dresselhaus spin-orbit 
coupling.
 Starting from the phenomenological Landau theory, we show that the long-wavelength dynamics of magnetization is governed by the Klein-Gordon equations. The standing-wave solutions of these equations describe \lq\lq particles\rq\rq\/ with effective masses, whose magnitudes and {\em signs} depend on the strength of the electron-electron interaction.
 The spectrum of the spin-chiral modes for arbitrary wavelengths is determined
 from the Dyson equation 
 for the interaction vertex.
 We propose  to observe spin-chiral modes via microwave absorption of standing waves confined by an in-plane profile of the spin-orbit splitting. 
\end{abstract}

\affiliation
{
Department of Physics, University of Florida, P. O. Box 118440,
Gainesville, FL 32611-8440}

\maketitle

\textit{Introduction.}---The rapidly developing field of spintronics aims to manipulate electron spins  
by electric rather than magnetic fields.
Since spin-orbit (SO) interaction allows for such a coupling, 
electron systems with SO interaction have been under intense study. 
A particularly interesting issue is the role of the electron-electron interaction in such systems \cite{raikh,saraga}. 
SO-coupled Fermi liquids (FLs) are expected to exhibit a rich variety of effects, which arise only from a combination of the electron-electron and SO interactions, such as spin-split and Rashba phases~\cite{wu,
hirsch}, unusual Friedel oscillations~\cite{badalyan,zak}, and spin textures~\cite{chesi}, to name just a few.  The focus of this 
Letter is 
on
the collective excitations in a SO-coupled FL.

The effect of the SO coupling on the electron spin can be thought of as resulting from an effective magnetic field which, in contrast to the real field, depends on the magnitude and direction of the electron momentum. With this analogy in mind, collective modes in 
an
SO-coupled FL are somewhat similar to spin waves in a FL subject to a (real) magnetic field \cite{silin,he3,metals}.
Spin waves occur because the exchange interaction couples precessing spins located at some distance from each other; this results in a dispersive mode 
which starts off at the unrenormalized (thanks to the Kohn's theorem) Larmor frequency
 and decreases with the wavenumber.
In the case of an
SO-coupled Fermi gas, the components of the Kramers doublet are split even in the absence of the external magnetic field. The SO-split states differ by their 
chirality,
i.e., 
a correlation in the directions of the electron momentum and spin.
The rate of direct transitions between the chiral branches of the spectrum 
determines
the frequency of the (zero-field) {\em combined} \cite{Rashba} or {\em chiral} spin resonance \cite{Finkelstein}. In 
an
SO-coupled FL, 
SU(2) invariance of electron spins is broken; as a result, there is not one but at least two resonances at $q=0$, corresponding to excitations of the in- and out-plane electron spins  \cite{Finkelstein}. 

In this Letter, we predict a new type of collective modes in a 
two-dimensional (2D)
  FL with SO coupling: {\em chiral spin waves}. The macroscopic equations of motion for the modes are derived using the quantum Boltzmann equation and the phenomenological FL theory. 
In the limit of small $q$ and in the absence of damping, these equations assume a form of 
Klein-Gordon equations for 
the
in- and out-of-plane components of magnetization. The standing-wave solutions of these equations can be thought of massive \lq\lq particles\rq\rq\/ with effective masses that depend on the strength of the electron-electron interaction. These masses not only differ in magnitude but also may be of opposite {\em signs}.
The SO-splitting, $\Delta$, plays the role of a potential energy of these particles. A lateral modulation of $\Delta$ along a 2D electron 
 (2DEG) plane acts as a potential well 
 confines particles. We propose to observe standing spin-chiral waves via microwave absorption in the presence of a local gate voltage which modulates the SO-coupling.    

\begin{figure}
\includegraphics[scale=0.22]{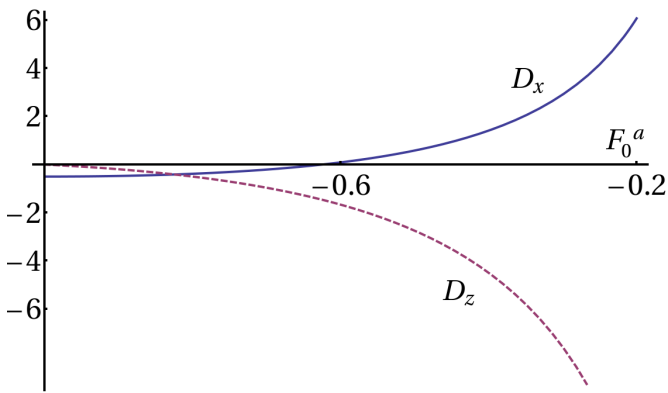}
\caption{(color on-line). Stiffnesses of the $x$-mode (solid) and $z$-mode (dashed) as a function of $F^a_0$.} 
\label{fig:dzdx}
\end{figure}

\textit{Equations of motion.}---We consider a 2D system of electrons in the presence of Rashba SO coupling ($\alpha$), described by 
the Hamiltonian \cite{Rashba}
$H=\frac{p^2}{2m}+\alpha\left(\bg{\sigma}\times\bom{p}\right)\cdot\hat{z}+H_{\mathrm{int}}$,
where $m$ is the effective electron mass, $\bg{\sigma}$ are the Pauli matrices, $\hat{z}$ is the unit vector along the normal to the 2DEG plane, and $H_{\mathrm{int}}$ 
entails
the electron-electron interaction.
We assume that the splitting of the Rashba subbands, $\Delta=2|\alpha|p_F$ (where $p_F$ is the Fermi momentum at $\alpha=0$), is much smaller than the Fermi energy. In this case, the SO coupling can be treated as a perturbation \cite{Finkelstein}.
A key quantity in 
the Landau's phenomenological theory of a Fermi liquid  
is the deviation of the occupation number matrix for quasi-particles (QPs), $\delta\hat{n}_{\bom{p}}(\bom{r},t)$, from its equilibrium value, $n^0$. The Boltzmann equation can be written as
\beq
 \partial_t\delta \hat{n}
 +i[\hat{n},\hat{\varepsilon}]_{-}+\bom{v}\cdot\bg{\nabla}_{\bom{r}}\delta \hat{n}
-\frac{1}{2}
\left[
\bg{\nabla}_{\bom{r}}\delta \hat{\varepsilon} ,\partial_{\bom{p}} n^0
\right]_{+}
= \left(\frac{\partial \delta \hat{n}}{\partial t}\right)_{\mathrm{coll}},
\label{1}
\eeq
where $\hat{\varepsilon}$ plays a role of the $2\times 2$ Hamiltonian for QPs and is a functional 
$\delta\hat{n}$
and $\left[A,B\right]_{\pm}$ denotes (anti)commutator of $A$ and $B$.  (For brevity, the dependences of $\delta{\hat n}$ on $\bom{p}$, $\bom{r}$, and $t$ are suppressed.)  The right-hand-side of Eq.~(\ref{1}) describes scattering of QPs, which we assume to be dominated by disorder.
Treating the SO coupling as a perturbation to the SU(2) symmetric FL, we follow the notations in \cite{Finkelstein} and represent $\delta \hat{n}$ as a sum of the perturbations due to the SO coupling and due to external forces
\beq \delta \hat{n}=\delta \hat{n}_{\mathrm{SO}}+\delta \hat{n}_{\mathrm{ext}} = 
\partial_{\varepsilon}n^0
\delta \hat{\varepsilon}_{\mathrm{SO}}+
\partial_{\varepsilon}n^0
\hat{u},
\label{deltan}
\eeq 
where 
$\hat{u}_{\bom{p}}(\bom{r},t)=u^i_{\bom{p}}{(\bom{r},t)} \tau^i$ , 
$\tau^1=-\sigma_z$ , $\tau^2=\bg{\sigma}\cdot \hat{p}$, $\tau^3=\left(\bg{\sigma}\times\hat{p}\right)\cdot \hat{z}$,
and $\hat{p}=\bom{p}/p$.
The
components of magnetization are expressed via $\hat{u}^i_{\bom{p}}$, projected onto the Fermi surface, 
as
\beq M_i=\frac{g\mu_B}4 \nu_F \int \frac{d\theta}{2\pi} \mathrm{Tr}\left(\sigma^i\hat{u}_{\bom{p_F}}\right),\eeq
where $g$ is the bare Land{\' e} factor of the electron, $\nu_F=m^*/\pi$ is the (renormalized) density of states, $\theta$ is the polar angle of $\bom{p}$, and $\mu_B$ is the Bohr magneton.
To exploit the in-plane symmetry, we set $M_y=0$
and keep only $M_x$ and $M_z$.
A deviation of the QP occupation number from the equilibrium results in a change of the QP energy
\bea \delta \hat{\varepsilon}& =&\delta \hat{\varepsilon}_{\tm{SO}} - \frac{\nu_F}{2} \int \frac{d\theta'}{2\pi} \
\mathrm{Tr '}\left({\hat f}_{\bom{p},\bom{p'}} \hat{u}'_{\bom{p}'}\right)\notag\\
 &&\delta \hat{\varepsilon}_{\tm{SO}}=\alpha^* p_F \tau^3=
 \Delta
 \tau^3/2,
\eea
where $\hat{f}_{\bom{p},\bom{p'}}$ is the Landau function, 
and prime refers to spin quantum numbers of the electron with momentum $\bom{p'}$. The effect of $\delta{\hat n}_{\tm{SO}}$ on $\delta{\hat\epsilon}$ is accounted 
for
via renormalization of the Rashba coupling  $\alpha\to\alpha^*=\alpha/(1+F^{a}_1/2)$, where $F^{a}_{\ell}$ is the $\ell$th harmonic of the spin part of the Landau function \cite{raikh,Finkelstein}.
To leading order in SO coupling, the collision integral due to short-range impurities can be written as $-(\hat{u}_{\bom{p}}-\langle\hat{u}\rangle)/\tau$
where $\langle\hat{u}\rangle$ is the average 
over the directions of the momentum
and $\tau$ is the 
impurity mean free 
time \cite{comment_imp}.
To the same accuracy, it suffices to keep the SU(2)-invariant form of the Landau function
\beq
\nu_F\hat{f}_{\bom{p},\bom{p'}}=F^s(\vartheta)\hat{I}\hat{I}'+F^a(\vartheta)\bg{\sigma}\cdot\bg{\sigma}',
\eeq
where $\vartheta$ is the angle between $\bom{p}$ and $\bom{p'}$ and both momenta are projected onto the Fermi surface.
We further adopt the $s$-wave approximation, in which $F^a=\tm{const}\equiv F^a_0$.  
This approximation allows one to obtain a closed-form solution of Eq.~(\ref{1}) without affecting the results qualitatively. 
With this assumption, one arrives at a closed system for $M_x$ and $M_z$:
\begin{subequations}
\bea
4M_x&=& 
F^a_0\int \frac{d\theta}{\pi}\left(\cos^2\theta\left(\Pi_{+-}+\Pi_{-+}\right)+ 2\sin^2\theta\Pi_{++} \right )M_x
\notag \\
&-&iF^a_0\int \frac{d\theta}{\pi}\cos\theta \left( \Pi_{+-}-\Pi_{-+}\right)M_z
\label{XZ} \\ 
2M_z&=&F_0^a\int\frac{d\theta}{2\pi}\left[\Pi_{\pm}\left(M_z+iM_x\right)+\Pi_{\mp}\left(M_z-iM_x\right)\right],
\label{XZ_1}
\eea
\end{subequations}
where $\Pi_{+-}=\left(\Delta+\bom{v}\cdot\bom{q}+\frac i {F^a_0\tau}\right)\left(\Omega-\Delta-\bom{v}\cdot\bom{q}\right)^{-1}$
with $\Omega=i{\partial_t}+i/\tau$, $\bom{q}=-i\bg{\nabla}_{\bom{r}}$,  and $\Pi_{-+}$ and $\Pi_{++}$ are obtained from $\Pi_{+-}$ by substituting $\Delta\rightarrow -\Delta$ and $\Delta=0$, respectively.
The denominators of $\Pi_{ss'}$ are inverse operators in space and time: keeping $M_i$ to the right of 
$\Pi_{ss'}$ emphasizes that. 
To obtain 
macroscopic equations of motion, we expand Eqs.~(\ref{XZ}) and (\ref{XZ_1}) to order $q^2$. 
In the ballistic limit ($\tau\rightarrow\infty$), the equations of motion are of the Klein-Gordon type:
\begin{subequations}
\bea
-{\partial_t}^2M_x&=&\Delta^2\left(1+\frac {F^a_0}{2}\right) M_x - 
D_x
\bg{\nabla}^2_{\bom{r}} M_x
\label{b}\\
-{\partial_t}^2M_z&=&\Delta^2 (1+F^a_0)M_z-
D_z,
\bg{\nabla}_{\bom{r}}^2 M_z,
\label{b_1}
\eea
\end{subequations}
where the mode stiffnesses depend on $F_0^a$ as 
\begin{subequations}
\bea 
 D_x
&=&-
\left[\frac{2}{F^a_0}+\frac {17} 4 +\frac {13} 8 F^a_0-\frac{(F^a_0)^2}{16(1+F^a_0/2)}\right]v_F^2
 \label{b1}\\
 D_z
&=&
\left[\frac{4}{F^a_0}+\frac {13} 2+ \frac 5 2 F^a_0\right]v_F^2.
\label{b1_1}
\eea 
 \end{subequations} 
Consequently, the dispersions of the modes are $\Omega_i^2=\Delta^2 (1+F^a_0\delta_i)+D_iq^2$, where $\delta_x=1/2$ and $\delta_z=1$.
At $q=0$, 
these
 equations reduce to  
 chiral spin resonances in the $s$-wave approximation \cite{Finkelstein}.
For a repulsive interaction,
 $F_0^a$ varies in between $0$ (free electrons)  and $-1$ (a ferromagnetic instability). While $D_z$
 is positive within this interval, $D_x$ changes sign at $F_0^a=F_c\approx-0.625$ (cf. Fig.~\ref{fig:dzdx}). 
For
 $F_c<F^a_0<0$, the signs of $D_x$ and $D_z$ are opposite.  
In the presence of damping, the $q=0$ form of Eqs.~(\ref{b}) and (\ref{b_1}) changes to
\begin{subequations} 
\bea
-\partial_t\left(\partial_t+\frac 1 {\tau}\!\right)^2 M_x&=&\Delta^2\left[\left(1+\frac{ F^a_0} {2}\right)\!\partial_t+\frac {1+F^a_0}{2\tau}\right]M_x.\notag\\
\label{c_2}
\\
-{\partial_t}\left(\partial_t+\frac 1 {\tau}\right)M_z&=&\Delta^2 (1+F^a_0)M_z.\label{c_1}
\eea
\end{subequations}
These equations describe Dyakonov-Perel 
spin relaxation
\cite{DP} 
renormalized by the electron-electron interaction.
The modes are well resolved
in the balistic limit, $\Delta\tau\gg 1$. 

\textit{Exact spectrum of the collective modes.}---  To study the spectrum of the collective modes for arbitrary $q$, we consider the Dyson equation for the  
scattering amplitude in the limit $\tau\to\infty$ \cite{AGD}
\bea
&\Gamma&_{s,r;s',r'}(P,K;Q)=\tilde{\Gamma}_{s,r;s',r'}(P,K
) \label{d1}
\\
&+&\int_{P'}\tilde{\Gamma}_{s,t;s',t'}(P,P'
)
\Phi_{tt'}(P';Q)\Gamma_{t',r;t,r'}(P',K;Q),\notag
\eea
where 
$\tilde{\Gamma}$ is the regular vertex, the
 \lq\lq four-momenta\rq\rq\/ are defined as $P=(\omega,\bom{p})$ etc., and 
 $s\dots t'=\pm 1$ label the Rashba subbands. The particle-hole correlators
 are given by
$\Phi_{ss'}(P,Q)=
(2\pi iZ^2/v_F)\delta(\omega)\delta(p-p_F)
\Pi_{ss'}$, with $\tau=\infty$ and  $\Omega\to\Omega+i0\tm{sgn}\Omega$ in
$\Pi_{ss'}$. Projecting $\Phi_{ss'}$ onto the Fermi surface in the absence of the SO coupling is permissible to leading order in $\Delta$; an explicit dependence on $\Delta$ is kept in $\Pi_{ss'}$.

To investigate the collective modes in the spin sector, we need to keep only the spin part 
of ${\tilde \Gamma}$ which, in the $s$-wave approximation, 
is identified as 
\beq Z^2\nu_F\tilde{\Gamma}^a_{s,t;s't'}(P,P')=F^a_0 
\langle s'\bom{p}\vert\bg{\sigma}\vert s\bom{p}\rangle
\cdot
\langle t'\bom{p}'\vert\bg{\sigma}\vert t\bom{p}'\rangle,
 \eeq
where $Z$ is the QP renormalization factor and $\langle s'\bom{p}'\vert\bg{\sigma}\vert s\bom{p}\rangle$ are the Pauli matrices in the chiral basis
\beq
\vert s,\bom{p}\rangle=\left(
\begin{array}{ccc}
1 \\ -ise^{i\theta}
\end{array} 
\right).
\eeq
Even in the $s$-wave approximation, $\tilde{\Gamma}^a_{s,t;s't'}$ depends on the directions of the electron momenta via $\bg{\sigma}_{ss'}$.
Since Eq.~(\ref{d1}) holds for any $K$, the 
vertex can be factorized as 
\beq\Gamma_{s,r;s',r'}(P,K;Q)=\eta_{ss'}(P;Q)\eta_{r,r'}(K;Q)
\eeq
Near the poles of $\Gamma$, we have  
\beq
\eta_{ss'}=
\frac{F^a_0}{2}
\sum_{t,t'}
\int \frac{d\theta'}{2\pi}
\langle s'\bom{p}_F\vert\bg{\sigma}\vert s\bom{p}_F\rangle\cdot
\langle t'\bom{p}'_F\vert\bg{\sigma}
\vert t\bom{p}'_F\rangle
\Pi_{tt'}\eta_{t't}.
\label{4}
\eeq

Changing the variables as $\mu_{ss'}=\Phi_{s's}\eta_{ss'}$ and expanding $\mu_{ss'}$ over a complete basis set as
$\mu_{ss'}=\sum_{n=0}^{\infty}\left( \mu_{ss'}^{,n}\cos{(n\theta)}+\bar{\mu}_{ss'}^{,n}\sin{(n\theta)}\right)$,
we cast Eq.~(\ref{4}) into the form of  Eqs.~(\ref{XZ}) and (\ref{XZ_1})
with 
$M_z\rightarrow{\mu}_{+-}^{,1}+{\mu}_{-+}^{,1}$ and
$M_x\rightarrow\left({\mu}_{+-}^{,1}-{\mu}_{-+}^{,1}\right)/2+i\bar{\mu}_{++}^{,1}.$

\begin{figure}
\includegraphics[scale=0.24
]{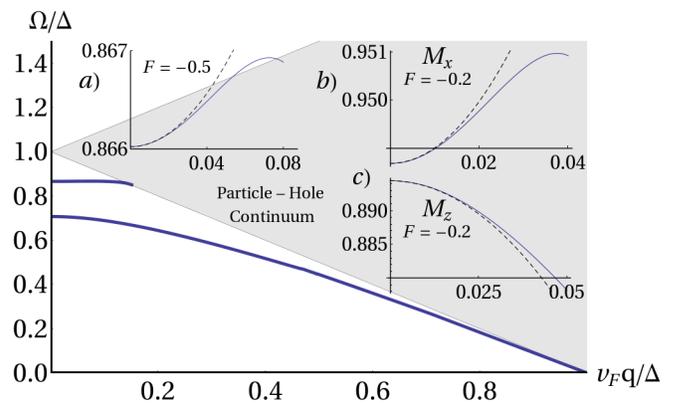}
\caption{(color on-line).
Spectrum of chiral spin waves for $F^a_0=-0.5$. 
Inset {\em a}): zoom of the small-$q$ region 
for the 
$x$-mode. Insets {\em b}) and {\em c}): 
spectra of the $x$- and $z$ modes
respectively, 
for $F^a_0=-0.2$ and small values of q. The dashed curves represent the parabolic approximation.
}
\label{fig:s}
\end{figure}

The resulting
angular 
 integrals can be solved for arbitrary values of $q$, after which the spectra of the modes 
 are
  found numerically.  
Figure ~\ref{fig:s} shows
the spectra
for $F
^a_0
=-0.5$. 
The higher-frequency mode is the spin-chiral wave of $M_x$, which runs into the 
particle-hole continuum
at $q<\Delta/v_F$. 
The lower-frequency mode is the spin-chiral wave of $M_z$
which merges with the continuum at $v_Fq=\Delta$.
Beyond the $s$-wave approximation,
the number of spin-chiral modes is infinite but the $q=0$ frequencies of the modes corresponding to higher harmonics 
are located closer the particle-hole continuum
and are thus damped heavier than the low-harmonic ones.

\textit{Experimental setup.}---
For standing-wave solutions,  
$M_i\sim \exp(i\Omega_it)$,
Eqs.~(\ref{b_1}) and (\ref{b}) are  
transformed into the 
\lq\lq Schroedinger  
equations\rq\rq\/ for massive particles 
\beq
\left[-\frac{1}{2\mathfrak{m}_i}\partial_{\bom{r}}^2 
+
V_i(\bom{r})\right]M_i=E_iM_i,
\label{Schroedinger}
\eeq
where $i=\left\{x,z\right\}$, the \lq\lq effective masses\rq\rq\/ are related to the stiffnesses in Eqs.~(\ref{b1}) and (\ref{b1_1}) via 
$\mathfrak{m}_i=1/2D_i$, $E_
i=\Omega_i^2$, and 
$V_
i(\bom{r})=\Delta^2(\bom{r})(1+F^a_0\delta_i
)$ 
 are the \lq\lq potential energies\rq\rq\/, which we now allow to vary slowly (compared to the electron wavelength) in the 2DEG plane. 
The lateral variation of $\Delta$ confines the spin-chiral modes and thus allows to extract the information about their dispersion, similar to how it was done for spin waves in He3
 \cite{he3} and alkaline metals \cite{metals}. 

\begin{figure}
\includegraphics[scale=0.3
]{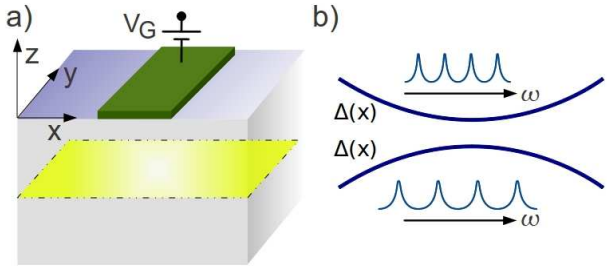}
\caption{\footnotesize 
(color on-line). 
a) Sketch of the suggested experimental setup.  
Top gate modulates the SO splitting. 
b) Top: 
A minimum in
$\Delta$
confines the $x$-mode with a 
positive
effective mass.
Bottom: 
A maximum in $\Delta$ 
confines the $z$-mode. 
The corresponding microwave absorption spectra are shown schematically in arbitrary units.} 
\label{setup}
\end{figure}

The effective mass of the $z$-mode is negative for any $F^a_0$ within the interval from $-1$ to $0$. Therefore, the $z$-mode
is confined by a {\em potential barrier} in $\Delta$, as shown in the bottom part of Fig~\ref{setup} $b$. The effective mass of the $x$-mode is positive for $F_c<F_0^a<0$ and negative for $-1<F_0^a<F_c$ ($F_c\approx -0.625$).
In the former case, the $x$-mode is confined by a potential well in $\Delta$, as shown in the top part of Fig~\ref{setup} $b$; in the latter case, the $x$-mode is confined in the same way as the $z$-mode, i.e., by a potential barrier. We propose to modulate $\Delta$ by applying a gate voltage to a part of the 2DEG. The width of the gate 
should be
chosen to be much larger than the electron wavelength, so that the electron motion 
would not be affected by the gate. Suppose that $F_c<F_0^a<0$, so that the effective masses of the $x$ and $z$ modes are of the opposite signs. In this case, a gate voltage of certain polarity confines only one type of modes. Discrete energy levels of the confined mode can be detected by microwave absorption. Although it is not {\em a priori} known which of the modes is confined, the control experiment would be to reverse the polarity of the gate voltage,
which would result in confining 
the mode with the opposite sign of the effective mass. Since not only the signs but also the magnitudes of the effective masses of the two modes are different, the distances between the peaks in the absorption spectra would change
on reversing the polarity of the gate voltage.
If $-1<F_0^a<F_c$,  both modes are either confined or deconfined for a given polarity of the gate voltage. 
By reversing the polarity, one would either suppress absorption or see a dense absorption spectrum.
The profile of $\Delta(\bom{r})$ must satisfy certain requirements. To be specific, we focus on the $z$-mode with the negative effective mass. Suppose that $\Delta$ varies along the $x$ axis in a stepwise manner, i.e., $\Delta(x)=\Delta_0=\tm{const}$ for
$|x|>a/2$ and $\Delta(x)=\Delta_0+W=\tm{const}$ for $|x|<a/2$ with $W>0$.  
A one-dimensional (symmetric) potential well has at least one bound state. 
However,
 to distinguish between single spin-chiral resonances, which exist even in the absence of the interaction,
and true quantized spin-chiral waves, one needs to observe several bound states. Using Eq.~(\ref{Schroedinger}), 
we find
the minimal condition for having more than one bound state 
as
$\left(\Delta_0W+W^2/2\right)\left(1+F_0^a\right)|\mathfrak{m}_z|a^2\geq 1$ \cite{LandauQM}, which implies
the ratio of $a$ to the SO length, $\lambda_{\tm{SO}}\equiv 1/2m|\alpha|$, should exceed a threshold value: \beq
\frac{a}{\lambda_{\tm{SO}}}\geq \left(2\frac{\left\vert\frac{4}{F^a_0}+\frac{13}{2}+\frac{5}{2}F^a_0\right\vert}{1+F_0^a}\right)^{1/2}\!\!\!\!\!\frac{1}{\left(W/\Delta_0+W^2/2\Delta^2_0\right)^{1/2}}.\label{cond}
\eeq
 In a GaAs heterostructure with $\alpha=5\; \tm{meV}\cdot\tm{\AA}$ \cite{miller:2003}, ${\lambda_{\tm{SO}}}\approx 1 \;\mu\tm{m}$.
According to Eq.~(\ref{cond}),  $a$ should be larger than $6.6\;\mu\tm{m}$ for $F_0^a=-0.3$ \cite{Foa} and  $W/\Delta_0=0.5$.
For larger $|F_0^a|$ and $W/\Delta_0$, the threshold value of $a$ is closer to ${\lambda_{\tm{SO}}}$. The condition
on the observation of the $x$-mode is more stringent, as this mode runs into the continuum at $q 
\approx
0.2/\lambda_{\tm{SO}}$ (cf. Fig.~\ref{fig:s}). Therefore, the $x$-mode is observable only for $a\gtrsim 30\lambda_{\tm{SO}}$.

 The second condition is that
the distance between the bound states must be larger than their width, which is of order $1/\tau$ in the ballistic regime.  For a potential well with a few bound states, this condition amounts to $|\mathfrak{m}_z|a^2\ll \tau^2$.
For $|F_0^a|\sim 1$, the last condition translates into $a\ll v_F\tau$, which is the same as the condition for the ballistic regime, i.e., $\Delta\tau\gg 1$.  Assuming that $\alpha$ does not depend on the number density $n$, we find that $\Delta\tau=3.5\times 10^{-6}\sqrt{n[10^{11}\mathrm{cm}^{-2}]}\mu[\mathrm{cm}^2/\tm{V}\tm{s}]$
 in a GaAs heterostructure, 
 where
 $\mu$ is the mobility. 
 The ballistic limit is achieved only if
$\mu 
>10^6\; \mathrm{cm}^2/Vs$.

An obvious way to excite the chiral spin modes is by the magnetic field, $\bom{B}$, oscillating near the resonance frequency.  Re-writing Eqs.~(\ref{b}) and (\ref{b_1}) as $\hat{L}_iM_i=0$, it is easy to see that in the presence of the field these equations become
$\hat{L}_i M_i=
g^2\mu_B^2 
\nu_F \Delta^2 \delta_iB_i/4
$. 
In addition,  
the SO interaction allows for a coupling of spins to an in-plane electric field, $\bom{E}$:
\begin{subequations}
\bea \hat{L}_x M_x&=&-\frac{g\mu_B}4 \nu_F \Delta^2 \frac{\alpha e
E_y}{\Omega_{x0}},\label{E_1}\\ 
\hat{L}_z M_z&=&4\left(1-\frac{F_0^a\Delta^2}{\Omega_{z0}^2}
\right)\left(\frac{\alpha e}{\Omega_{z0}}
|\bom{E}\times\hat{z}|\right)^2 M_z,
\label{E}
\eea
\end{subequations}
where $\Omega_{0i}$ are the resonance frequencies 
at $q=0$.
While the $x$-mode 
couples linearly   
to the electric field \cite{Finkelstein},
the   
$z$-mode is generated to second order in $E$. 
Equation (\ref{E}) describes a parametric resonance in $M_z$ excited by the electric field with frequency $\Omega_{z0}$.
The initial amplitude of $M_z$ can be provided by a pulse in $B_z$.

 It is worth noting that all of the results presented above remain the same if the Rashba SO interaction is replaced by the Dresselhaus one. If the Rashba and Dresselhaus interactions are present simultaneously, spin-chiral modes become non-sinusoidal. 
 Hence, it is better to perform the experiment on a symmetric quantum well which has only  the Dresselhaus but no Rashba interaction. 
 
We are grateful to K. Ensslin, Y. Lee, D. Loss, C. Marcus, A. Meyerovich, E. Rashba, S. Tarucha, and D. Zumb{\"u}hl for stimulating discussions.  The work was supported by NSF-DMR  0908029. D.L.M. acknowledges the support from the  Swiss NSF
\lq\lq QC2 Visitor Program\rq\rq\/ at  the University of Basel.


\begin{thebibliography}{99}
\bibitem{comment_imp} For finite-range impurities, the spin-relaxation rate is determined by the first 
three
harmonics 
of the correlation function of disorder \cite{Finkelstein}.  This circumstance does not affect the results qualitatively.
\end{thebibliography}
\end{document}